\documentclass[preprint,12pt]{elsarticle}

\usepackage{amssymb}
\usepackage{amsmath}
\usepackage{subcaption}
\usepackage{graphicx}

\begin{document}

\begin{frontmatter}



\title{Forward-backward correlations: A probe to study dynamical fluctuations}




\author{Somen Gope$^{*}$}
\author{Supriya Das}
\author{Saikat Biswas}
\address{Department of Physical Sciences, Bose Institute Unified Academic Campus, Kolkata, India \\
$^{*}$somengope30@jcbose.ac.in}

\begin{abstract}
The source of the fluctuations in the final state particles is the initial event-by-event fluctuation in energy density. Forward-Backward (F-B) correlation is one of the important probes to study such fluctuations. The results of F-B correlation are available in a wide range of energy, from STAR of RHIC to ALICE of LHC. It will be more interesting to study the dynamical fluctuation using F-B correlation at SIS100 energy too. In this study, an attempt has been made to investigate the F-B correlation in UrQMD-hydro generated data for 10 AGeV Au+Au collisions.


\end{abstract}




\end{frontmatter}

\section{Introduction}
One of the prime objectives of ultra-relativistic heavy ion collisions (HIC) is to find the experimental evidence of the deconfinement phase transition from hadronic matter to quark-gluon plasma (QGP). There are various hadronic models that are used to study and investigate the phase transition (PT) and fluctuations in the final state particles. The source of the fluctuations is the initial event-by-event fluctuation in energy density. There are several probes to study fluctuations in HIC. One such probe is the forward-backward (F-B) correlations. Short-range correlation (SRC) and long-range correlation (LRC) have been observed in high-energy heavy-ion collisions. Only SRC is observed in all energy ranges of collision only if two-particle measurement is considered. However, multiparticle production has never demonstrated SRC as the sole source of correlation. On the other hand, LRC is energy and system dependent. It has been reported that the strong correlation is observed in the production of charged particles in the region of mid-rapidity ($y=0$) with the particles emitted in the forward region of rapidity space \cite{paper001,paper002}. 

The hydrodynamic calculation of heavy-ion collision is found to be more realistic in explaining the properties of matter produced. The space-time evolution of the fireball can be explained through various observables, such as elliptic flow ($v_{2}$), harmonic flow ($v_{n}$), transverse momentum ($p_{T}$). The existence of dynamical fluctuation in hydrodynamic model-generated data is very much important to study. A brief study has already been reported in ref. \cite{paper3}. 

A rigorous study on F-B correlations in nucleus-nucleus collisions has been reported by NA22 \cite{paper01}, STAR \cite{paper02}, and ALICE  \cite{paper002} Collaborations. The same study would be significant to be carried out at the SIS100 energies. According to the hydrodynamical calculations, the deconfined phase boundary will be accessible at the highest SIS100 energy. In this article, an attempt has been made to study the correlation using the data generated by UrQMD, UrQMD-hydro, and UrQMD-hydro with disabling the Cooper-Frye approach.

\section{Mathematical Formalism}

Mathematical representation of F-B correlation is presented by correlation coefficient or strength of correlation ($b_{corr}$). Before defining $b_{corr}$, a few parameters are also essential to understand. Let ${\delta y}$ be the rapidity window that is located symmetrically from $y=0$ at distances $\pm y$ of the rapidity distribution of the charged particles. A pictorial view is shown in Fig. 1.

\begin{figure}[h!]
	\begin{center}
		\includegraphics[scale=0.7]{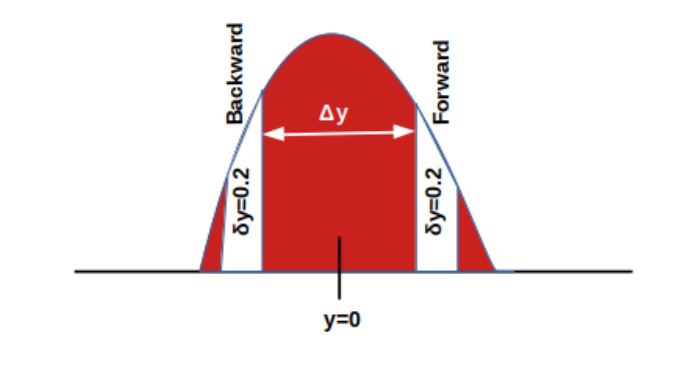}
		\caption{\label{fig1} (Color online) the cartoon illustrates the picture of forward = (${\Delta y}/2$, ${\Delta y}/2 + {\delta y}$) and backward = (${\Delta y}/2 - {\delta y}$, ${- \Delta y}/2$) rapidity regions.}\label{fig1}
	\end{center}
\end{figure}


The exact separation between the nearer ends of the forward and backward windows is therefore $2y$, which we treat as the F-B rapidity gap ($y_{gap}$). Let consider the first moment of the multiplicity in the forward (backward) window is $\langle n_{B} \rangle_{n_{F}}$. The strength of F-B correlation strength is mathematically defined as:

\begin{equation}
\langle n_{B} \rangle_{n_{F}} = a + b_{corr} \times \langle n_{F} \rangle 
\end{equation}

where the parameters $a$ \& $\langle n_{F} \rangle$ are the number of uncorrelated particles and the average number of particles in the forward rapidity region. The values of the correlation strength parameter ($b_{corr}$) can be positive or negative. The ($b_{corr}$) value varies with the types of correlation. One is short-range correlation (SRC), and another is long-range correlation (LRC). $b_{corr}$ vanishes for uncorrelated emission of particles. The F-B correlation strength can also be expressed by the ratio of the F-B multiplicity covariance and the variance of forward (backward) multiplicity as \cite{paper002,paper04,paper3}

\begin{equation}
b_{corr} = \frac{\langle n_{F}n_{B} \rangle - \langle n_{F}\rangle \langle n_{B} \rangle }{\langle n_{F}^2 \rangle - \langle n_{F} \rangle ^2} 
\end{equation}

where: $V_{FB} = \langle n_{F}n_{B} \rangle - \langle n_{F}\rangle \langle n_{B} \rangle$ \& 
$V_{FF}= \langle n_{F}^2 \rangle - \langle n_{F} \rangle ^2$

Also $b_{corr}$ is related to coefficient $\delta$ via following formula \cite{paper3}:

\begin{equation}
b_{corr} \propto e^{({-y_{gap}}^{2}/\delta^{2})}
\end{equation}

The coefficient $\delta$ is related to the correlation length ($\lambda$)
via \cite{paper3}:


\begin{equation}
\lambda = 2 \delta / \sqrt{\pi}
\end{equation}

Also, the event-by-event multiplicity fluctuation is calculated by measuring the variance and is given by

\begin{equation}
\sigma^{2} = \frac{V_{FF}^2 + V_{BB}^2 - 2 V_{FB}^2 }{\langle n_{F} + n_{B} \rangle} 
\end{equation}

\section{UrQMD $\&$ UrQMD-hydro model} \label{det_con}

The Ultrarelativistic Quantum Molecular Dynamics (UrQMD) model is a microscopic transport model \cite{paperM0}. It is used to successfully simulate the heavy-ion collisions \cite{paperM01}. In the UrQMD model, hadronic interactions are simulated using classical transport equations. The UrQMD uses the principle of detailed balance based on experimental \& theoretical calculation for getting the geometrical cross section of interactions. UrQMD is less effective to describe late-stage collision or interaction because of its microscopic nature. The late stage can be explained successfully by considering the fluid nature \cite{paperM1, paperM2, paperM3}. 

The UrQMD-hydro model is the micro+macro approach of UrQMD. It describes the hydrodynamic description of the fireball. The model switches to hydrodynamic behavior with preferred eqaution of state (EoS) when the collision system is in its final stages. In the final stages the Cooper-Frye equation is used to describe the transition from hydrodynamic degrees of freedom to particle degrees of freedom and then the particle information is transferred back to the microscopic UrQMD transport model. In the final stage, new particle formation is performed through hadronic cascade process. A more detailed description of the mentioned model and parameters related to this model are described in \cite{paperM4}.

In this article, the results from the UrQMD+Hydro (with Chiral EoS) model are compared to the default transport calculation (UrQMD-default). It is also noted that, the parameters (centrality, collision energy, collision system etc.) are same for diffrent modes of data.

\section{Results}
\vspace{-0.25 cm}
In the present study, three sets of UrQMD and UrQMD-hydro data have been generated for Au+Au collisions at 10 AGeV beam energy. 

\begin{figure}[htb!]
	\begin{center}
		\includegraphics[scale=0.5]{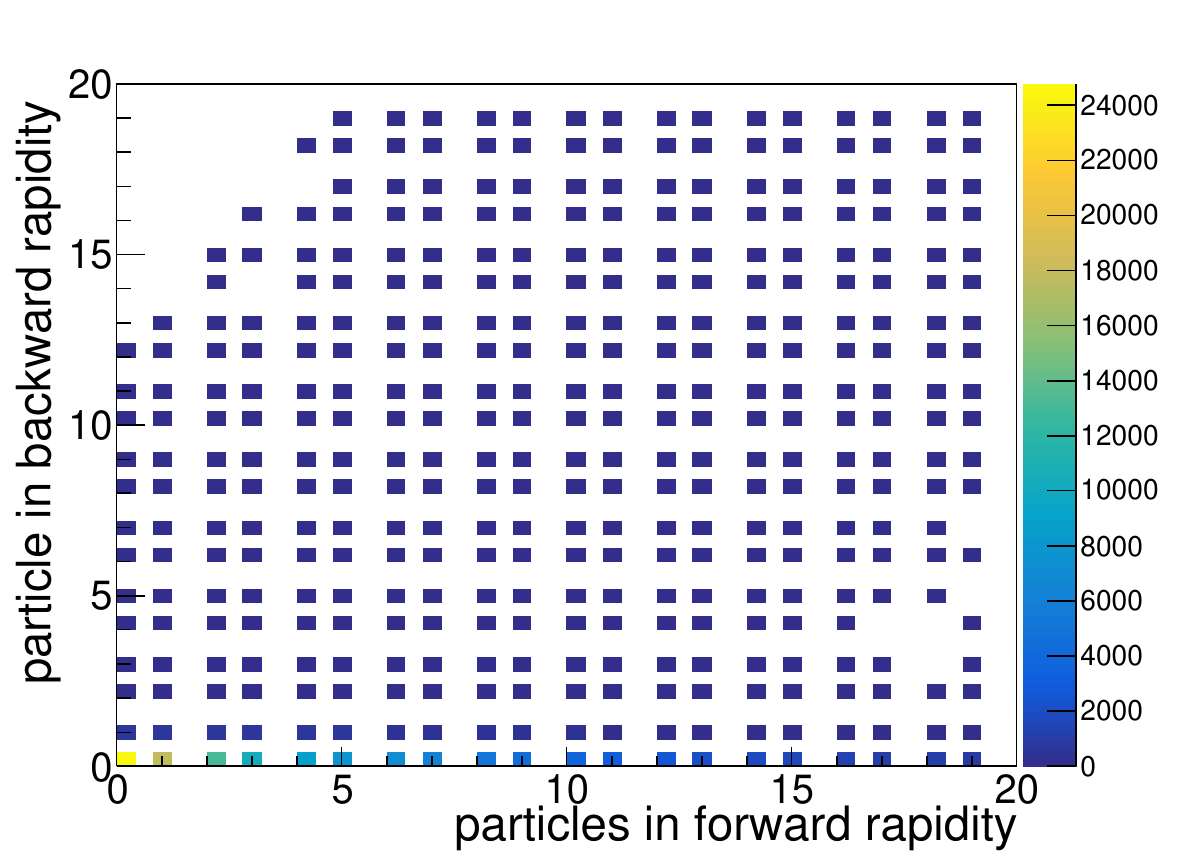}
		\caption{\label{fig1} 2D plot of distribution of particles in forward \& backward rapidity spaces.}\label{fig1}
	\end{center}
\end{figure}

The event statistics are 150 K, 100 K, \& 60 K for UrQMD-default, UrQMD-hydro, \& UrQMD-hydro with disabling the C-F approach, respectively. In this study, only minimum bias Au+Au data have been considered for the analysis.



A 2D plot drawn for the particles emitted in forward and backward rapidity regions is shown in Fig. 2. Figure shows that the number of particles in the forward rapidity region increases with the increase of the particles in the backward rapidity region. From this observation, one can say that there is a correlation between the particles emitted in either rapidity spaces.

\begin{figure}[h!]
	\begin{center}
		\includegraphics[scale=0.5]{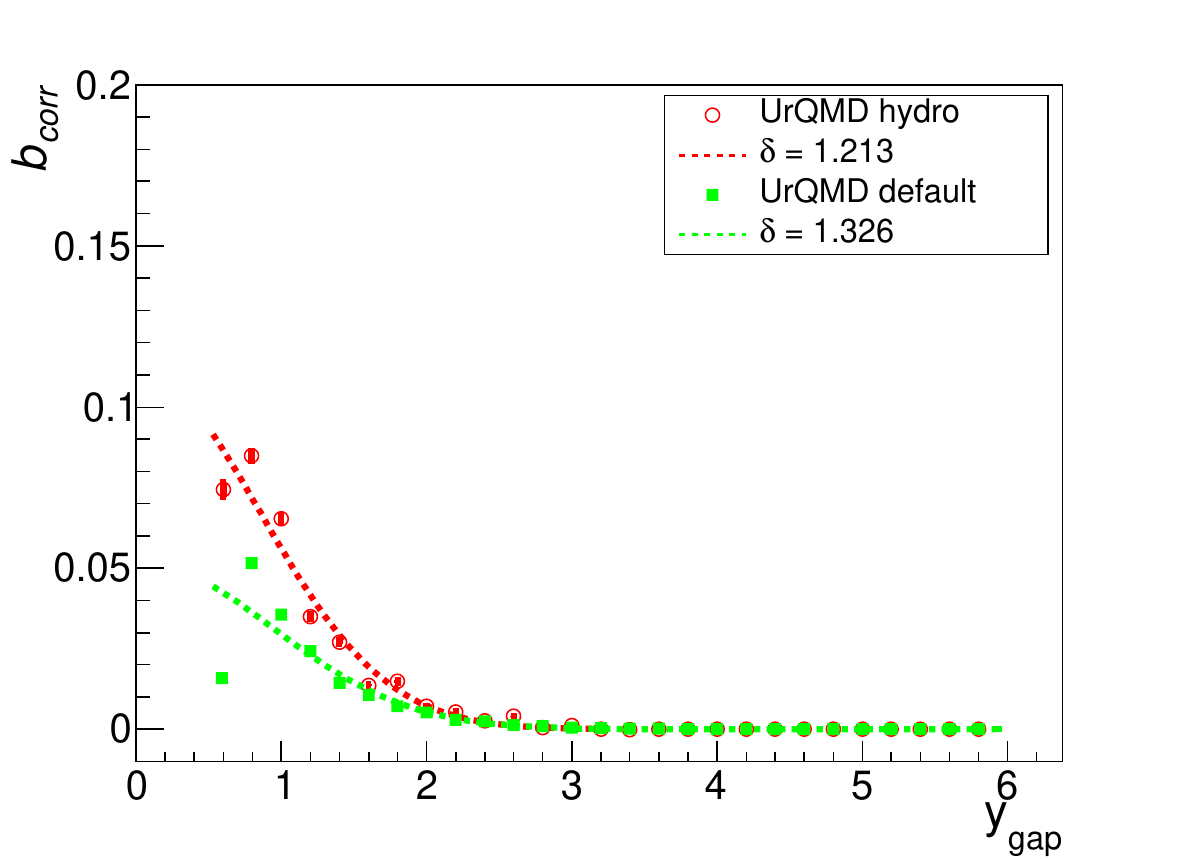}
		\caption{\label{fig1} The plot of variation of correlation coefficient with the rapidity gap for UrQMD-defaut \& UrQMD-hydro generated data.}\label{fig1}
	\end{center}
\end{figure}

Fig. 3 is the plot of correlation coefficient as a function of $y_{gap}$. From this plot, it is clearly seen that the value of $b_{corr}$ is maximum when the $y_{gap}$ is minimum or $y_{gap} \sim 0$. $b_{corr}$ decreases monotonically with the increase of the value of $y_{gap}$. The same trend has been observed for both sets (UrQMD-default \& UrQMD-hydro) of data. But $b_{corr}$ values are higher for UrQMD-hydro-generated data than that of UrQMD-default data.

The plots are fitted with the equation 3. The fitted parameter $\delta$ is related to correlation length ($\lambda$) by the equation 4. The values of correlation length and $\delta$ are listed in table 1.

\begin{table}[h!]
	\centering
	\begin{tabular}{l c c} 
		\hline
		Data type & $\delta$ & Correlation length ($\lambda$) \\ 
		\hline
		UrQMD default & 	1.326 $\pm$ 0.005 & 1.497 $\pm$ 0.006 \\ 
		Hydro default &  1.192 $\pm$ 0.004 & 1.345 $\pm$ 0.005 \\
		\hline
	\end{tabular}
	\caption{Table for the values of correlation length for different sets of UrQMD \& UrQMD hydro data.}
	\label{Table1}
\end{table}

\begin{figure}[h!]
	\begin{center}
		\includegraphics[scale=0.5]{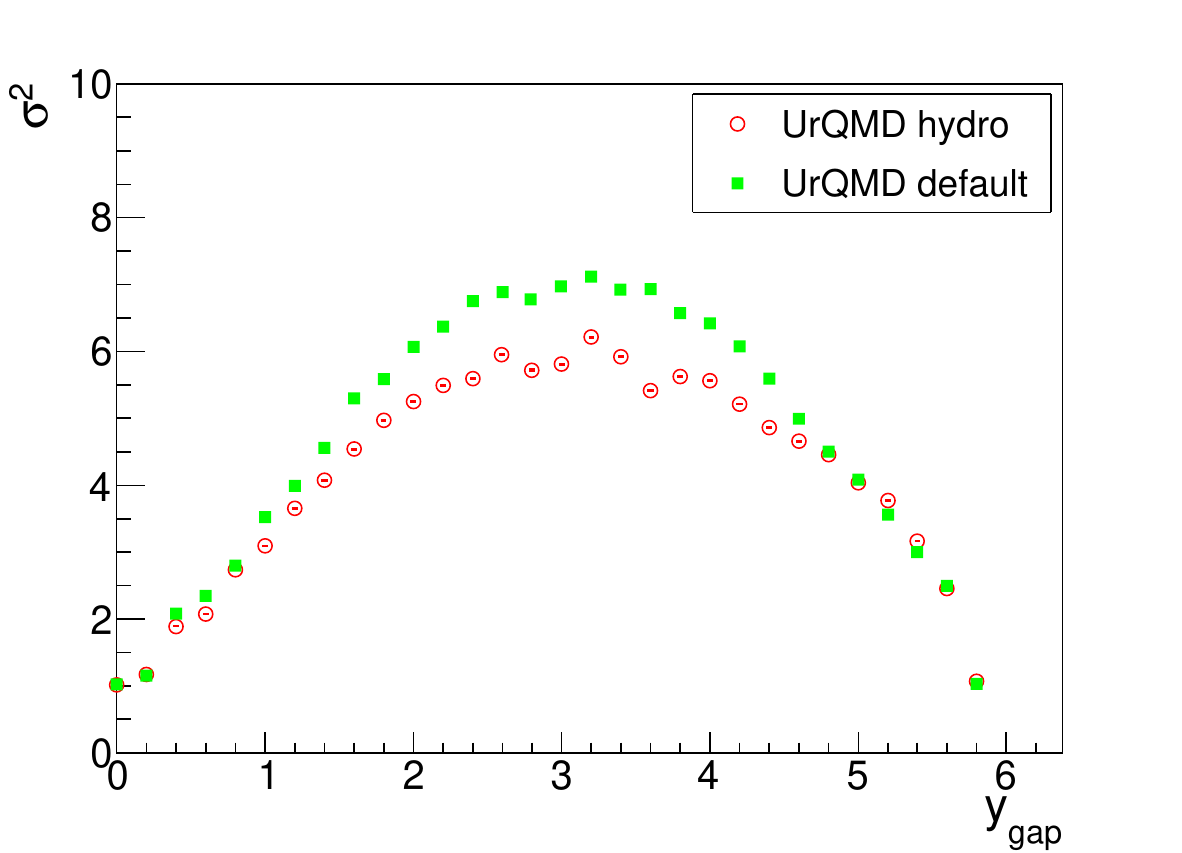}
		\caption{\label{fig1} $\sigma^{2}$ distribution for UrQMD-default \& UrQMD-hydro generated data.}\label{fig1}
	\end{center}
\end{figure}

\begin{figure}[htb!]
	\begin{center}
		\includegraphics[scale=0.5]{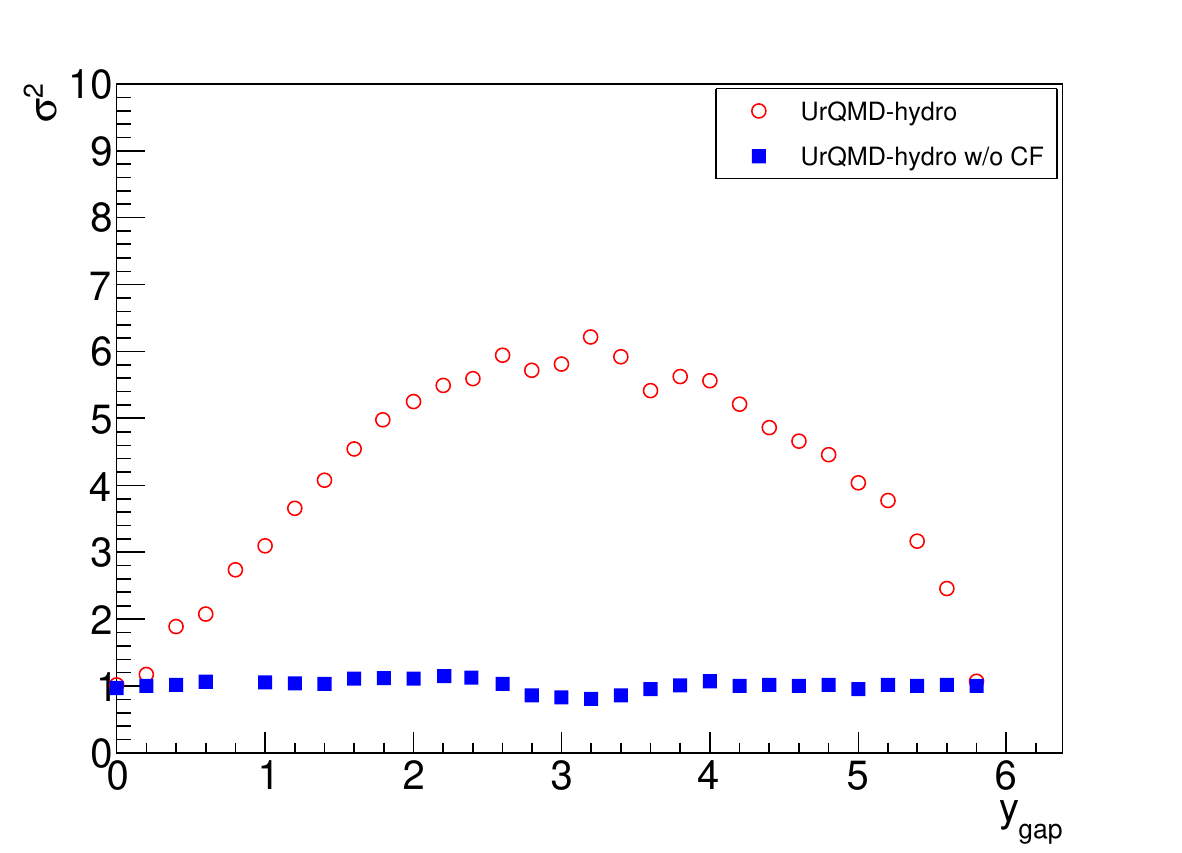}
		\caption{\label{fig1} $\sigma^{2}$ distribution for the data of UrQMD-hydro \& UrQMD-hydro with out CF.}\label{fig1}
	\end{center}
\end{figure}

However, $\sigma^{2}$ vs. $y_{gap}$ has been plotted to check the type of correlation in the particles produced in heavy ion collision. Fig. 4 shows the variation of $\sigma^{2}$ as a function of $y_{gap}$ for both the sets of data. It has been observed that for both the sets of data, the $\sigma^{2}$ value is greater than unity in all $y_{gap}$ values, that means LRC is the dominant.

Another set of data has been generated to check the effect Cooper-Frye approch in the result of correlation studies. The same plot of $\sigma^{2}$ vs $y_{gap}$ has been plotted using UrQMD-hydro generated data with disabling the C-F approach.  It is observed that the new data set shows no dependency with $y_{gap}$, indicating lack of correlation.

\section{Summary}

In this work, a systematic study on forward-backward multiplicity correlations for Au+Au system has been performed using various modes of UrQMD generated data. It is observed that the value of F-B correlation strength ($b_{corr}$) decreases as a function of $y_{gap}$.

The correlation length $\lambda$ is calculated from the value of $\delta$ and the values are greater than unity for both the sets of data. This is the signature of existence of long-range correlation (LRC). That also means that the particles are still correleted at the value of rapidity gap more than one unit.

The $\sigma^{2}$ value is also calculated for three sets of data and plotted as a function of $y_{gap}$. The value of $\sigma^{2}$ is greater than the Poisson limit (that means greater than 1) for the data sets UrQMD-default \& UrQMD-hydro default. But the data UrQMD-hydro with the disabling CF approach shows the flat behaviour of $\sigma^{2}$ as function of $y_{gap}$.  

\section*{Acknowledgements}
The authors would like to thank the UrQMD group for developing the model and provide the source code publicly.  Authors also would like to thank Mr. Subir Mandal for helping to generate the monte carlo data. Author also would like to thank DST and DAE, Govt. of India for providing the financial support through the project bearing number: SR/MF/PS-02/2010 (E-6133) dt 08/10/2021.

\end{document}